\documentclass[twocolumn,10pt]{article} 

\usepackage[square,numbers,sort&compress,comma]{natbib}

\usepackage{amsmath}
\usepackage{amssymb}
\usepackage{caption}
\usepackage{graphicx}
\usepackage{latexsym}
\usepackage{times}
\usepackage{adjustbox}
\usepackage{xcolor}
\usepackage{soul}
\usepackage[normalem]{ulem}
\usepackage{xcolor}
\usepackage{url}
	
\usepackage{color, colortbl}

\topmargin - 12pt 
\renewenvironment{abstract}%
              {
               \small
               {\bfseries \abstractname}
               \par
               \vspace{10pt}
              }

\renewcommand\abstractname{Abstract}

\newcommand{\nomenclature}
              [1]
              {
               \bgroup
               \flushleft
               \small\bf
               #1
               \par
               \egroup
              }

\renewcommand{\section}
              [1]
              {
               \bgroup
               \flushleft
               \small\bf
               \stepcounter{section}
               \arabic{section}. #1
               \par
               \egroup
              }

\renewcommand{\subsection}
              [1]
              {
               \bgroup
               \flushleft
               \small\em
               \stepcounter{subsection}
               \arabic{section}.
               \arabic{subsection}. #1
               \par
               \egroup
              }

\renewcommand{\subsubsection}
              [1]
              {
               \bgroup
               \flushleft
               \small\em
               \stepcounter{subsubsection}
               \arabic{section}.
               \arabic{subsection}.
               \arabic{subsubsection}. #1
               \par
               \egroup
              }

  \newcommand{\acknowledgement}
              [1]
              {
               \bgroup
               \flushleft
               \small\bf
               #1
               \par
               \egroup
              }

  \newcommand{\sectionbib}
              [1]
              {
               \bgroup
               \flushleft
               \small\bf
               #1
               \par
               \egroup
              }

\begin{document}

\title{\LARGE Critical diffraction of irregular structure detonations and their predictability from experimentally obtained $D$-$\kappa$ data\\
                  }

\author{{\large Farzane Zangene$^{a,*}$, Qiang Xiao$^{a, b}$, Matei Radulescu$^{a}$}\\[10pt]
        {\footnotesize \em $^a$Department of Mechanical Engineering, University of Ottawa
        , Ottawa, ON K1N6N5, Canada}\\[-5pt]
        {\footnotesize \em $^b$now at the National Key Laboratory of Transient Physics, Nanjing University of Science and Technology, Nanjing 210094, China}\\[-5pt]}
\date{}


\small
\baselineskip 10pt


\twocolumn[\begin{@twocolumnfalse}
\vspace{50pt}
\maketitle
\vspace{40pt}
\rule{\textwidth}{0.5pt}
\begin{abstract} 
The present work reports new experiments of detonation diffraction in a 2D channel configuration in stoichiometric mixtures of ethylene, ethane, and methane with oxygen as oxidizer.  The flow field details are obtained using high-speed schlieren near the critical conditions of diffraction.  The critical initial pressure for successful diffraction is reported for the ethylene, ethane and methane mixtures.  The flow field details revealed that the lateral portion of the wave results in a zone of quenched ignition.  The dynamics of the laterally diffracting shock front are found in good agreement with the recent model developed by Radulescu et al. (Physics of Fluids 2021).  The model provides noticeable improvement over the local models using Whitham's characteristic rule and Wescott, Bdzil and Stewart's model for weakly curved reactive shocks.  These models provide a link between the critical channel height and the critical wave curvature. The critical channel heights and global curvatures are found in very good agreement with the critical curvatures measured independently by Xiao and Radulescu (Combust. Flame 2020) in quasi-steady experiments in exponential horns for three mixtures tested.  Furthermore, critical curvature data obtained by others in the literature was found to provide a good prediction of critical diffraction in 2D.
These findings suggest that the critical diffraction of unstable detonations may be well predicted by a model based on the maximum curvature of the detonation front, where the latter is to be measured experimentally and account for the role of the cellular structure in the burning mechanism.  This finding provides support to the view that models for unstable detonations at a meso-scale larger than the cell size, i.e., \textit{hydrodynamic average} models, are meaningful.        

\end{abstract}
\vspace{10pt}
\parbox{1.0\textwidth}{\footnotesize {\em Keywords:} Diffraction; Irregular Structure Detonations; Critical Curvature}
\rule{\textwidth}{0.5pt}
\vspace{10pt}
\end{@twocolumnfalse}] 


\clearpage
\section{Introduction} \addvspace{10pt}
The diffraction of detonations at an abrupt area change is a fundamental problem of detonation dynamics. The critical tube diameter $d_{c}$ (or channel slot height $W_{c}$ in two-dimensional channels) is considered a measure of the detonability of detonations. It is known empirically to depend on other such dynamic parameters, offering an alternative measure of a length scale proportional to the detonation reaction zone thickness (cell size, hydrodynamic thickness, critical explosion length for direct initiation) \cite{Lee:1984, Lee:2008}. For a wide range of mixtures, empirical evidence suggests that $d_c/\lambda \simeq 13$ ($\lambda$, detonation cell size) for detonation diffraction from tubes, while a generally larger value is observed in detonations with very regular structures \cite{moen1986influence}.  The good correlation with cell size suggests the importance of the cellular dynamics in controlling the detonability of gases.  

\begin{figure}
\centering
\includegraphics[width=\columnwidth]{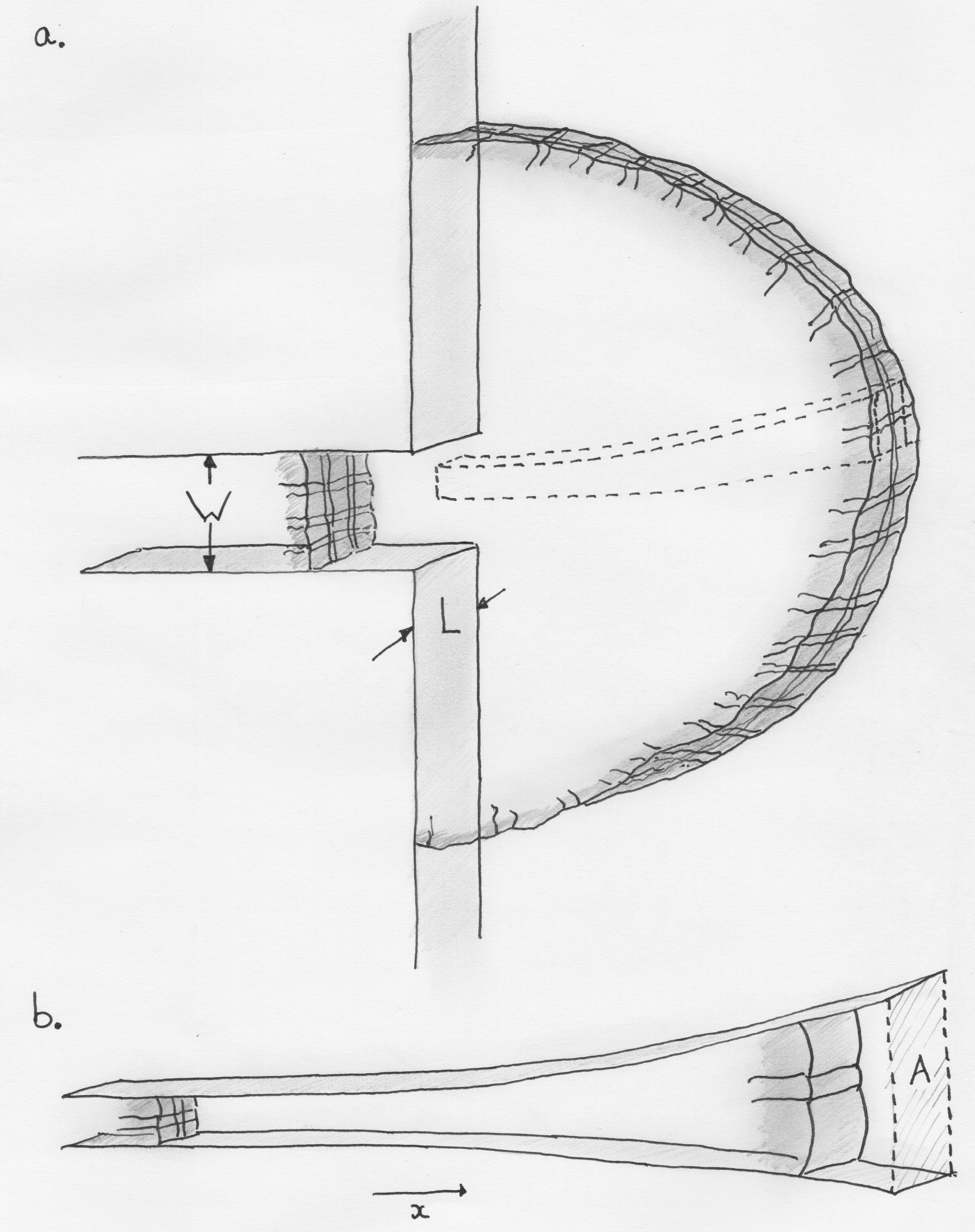}
\vspace{10 pt}
\caption{The diffraction of a detonation wave from a slot (a) and the concept of a unit-cell ray tube of lateral divergence rate $\kappa=\mathrm{d} \ln A / \mathrm{d}x$ to characterize the dependence $D(\kappa)$ for cellular detonations \cite{Radulescu1:2018, xiao:2020dynamics} .}
\label{fig:sketch}
\end{figure}

 For the equivalent problem in 2D, the situation is more complicated.  When a planar cellular detonation propagates in a channel of depth $L$ and height $W$ and transits into a much larger channel of the same depth $L$ (Fig.\ \ref{fig:sketch}), the empirical evidence suggests the limit $W_c/\lambda$ is very strongly influenced by the width of the channel $L$ \cite{thomas1986detonation, benedick1983large, meredith2010detonation, liu1984effect}.  For thin channels, where there are nominally less than one cell across, i.e., $L/\lambda \lesssim 1$, and the cellular structure approaches a 2D structure, the limit observed was $W_c/\lambda \simeq 10$ \cite{mitrofanov1964multifront, thomas1986detonation, edwards1979diffraction}.  In much wider channels in which the detonation cellular structure is 3D, the limit varies between $W_c/\lambda \approx3$  and $W_c/\lambda \approx6$, with a smaller value in more irregular systems.  At present, this variation from 3 to 10 is not well understood. 
 
A possible explanation is the known intrinsic difference in cellular dynamics between 2D and 3D cellular structures, preliminary results being reported by Crane et al.\ in these proceedings \cite{crane2022proci}; how a 3D cellular detonation may offer a more robust propagation mechanism remains to be explained, although multiple mechanisms are possible (Kolmogorov cascade only in 3D, differences between 2D and 3D shock dynamics and ensuing reactivity, etc...).  Also compatible with the lower limit of $W_c/\lambda \approx3$ in more irregular mixtures is the role of stochasticity.  Since the critical transmission always displays some level of stochasticity \cite{loiseau:2007} inherent from the cellular dynamics and manifested through discrete explosion centers, it is expected that wider channels favour transmission, as the probability of a single re-initiation center across the whole surface of diffracting shock increases proportionally with the channel width.  A single remaining explosion center could then\textit{ percolate} across the entire front, a scenario suggested to explain analogous results for detonation quenching in non-ideal energetic materials subjected to lateral flow divergence \cite{higgins2009measurement}.

The resulting anomalous scaling between the critical tube diameter and channel width $d_c/W_c$ has been recognized as a signature of the cellular structure and local stochasticity.  It is expected to be 2 based on simple geometric arguments relating the radius of curvature of the front and the characteristic scale controlling the diffraction $d_c$ or $W_c$ \cite{liu1984effect}.  The complications accounted above did not permit to warrant conclusive comparisons between experiments and models, except in very regular systems.  The present experiments in a 2D geometry address the limits in 2D. We conduct experiments in sufficiently wide channels in order to allow a 3D cellular structure, but sufficiently narrow to prevent the stochasticity associated with very wide channels, which is left for future study. 

\begin{figure*}
\centering
\includegraphics[scale=0.9]{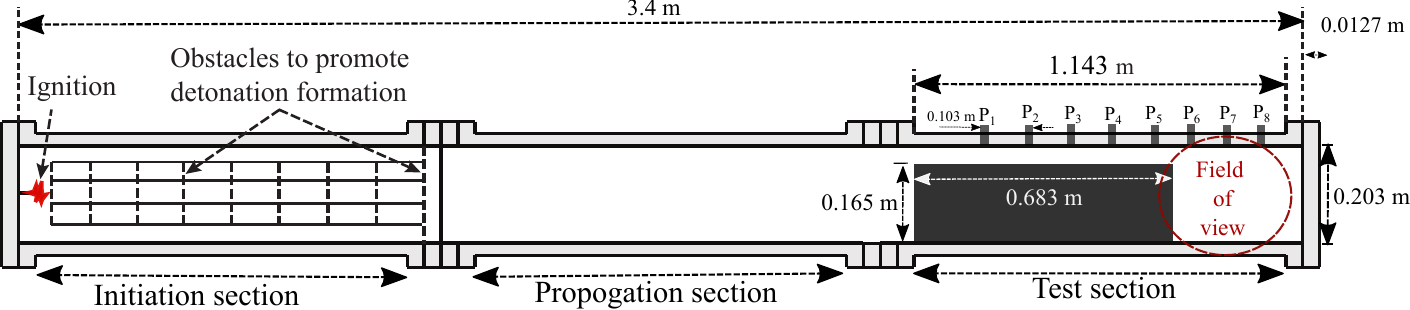}
\vspace{10 pt}
\caption{Schematic of the shock tube used in the diffraction experiments.}
\label{STDI}
\end{figure*}

The prediction of detonation diffraction from first principles, using detailed hydrodynamic simulations, is usually restricted to hydrogen detonations at low pressures, mixtures that are not overly sensitive to gas dynamics perturbations and have a very regular cellular structure \cite{Mevel:2017}. In most other detonations, simulations have been restricted to qualitative observations using model parameters in a global one-step description \cite{arienti:2005, Shi:2020}, sometimes far removed from the real thermo-kinetic data. These types of simulations also face a plethora of numerical and modelling difficulties, stemming primarily from the necessity to tackle small-scale diffusive phenomena in real detonations \cite{Radulescu:2007, Radulescu1:2018}. For these unstable detonations, the prediction of the critical conditions for diffraction relies on correlations of the type discussed above.

Recently, Nakayama et al.\ \cite{nakayama2013front} have demonstrated that dynamics of detonations in several mixtures can be approximately described by a unique relation between the detonation speed, $D$, and the mean curvature of the front, $\kappa$. These observations were borne out from experiments of detonation diffraction in curved channels with varying wave curvatures and wave speeds. While possible non-steady effects cannot be ruled out from these experiments, the good collapse of the entire data in terms of $D(\kappa)$ only suggests these be of minor importance. These observations strongly suggest that a Detonation Shock Dynamics approach, developed by Bdzil and Stewart \cite{bdzil:1989} and used in condensed phase detonation engineering applications \cite{lambert:2006}, can be a useful tool for predicting gaseous detonations in complex geometries.  In this approach, the dynamics of a cellular front at large scales could be modeled as a collection of individual mean ray tubes containing at least one cell (see Fig.\ \ref{fig:sketch}).   A simple experiment was recently devised by Radulescu and his students to measure the response of a unit cell to a global geometrical effect by using an exponential horn \cite{Radulescu:2018, xiao:2020dynamics, Xiao:2020} as to keep the mean curvature of the front $\kappa=\mathrm{d}\ln A/\mathrm{d}x$ constant, yielding the desired empirical shock speed curvature $D(\kappa)$ relation for closure.  This method thus permits to unambiguously determine the $D(\kappa)$ curves for a given explosive mixture at a hydrodynamic average scale larger than the cell size, as it automatically takes into consideration the complex cellular dynamics of detonations.  It also provides an estimate of the maximum curvature obtainable in quasi-steady state.

Recently, we have shown that the $D(\kappa)$ dependence obtained from the ZND model was found in reasonable agreement with experiment in very regular H$_2$/O$_2$/Ar detonations.   With a shock diffraction model linking the shock curvature on the axis to the channel width $W$, this permitted to reasonably predict the critical diffraction slot width $W$ in these systems \cite{Radulescu:2021, xiao2022ray}.  While the ZND model is not expected to work in more irregular detonations \cite{Xiao:2020}, we wish to determine whether this critical diffraction slot width $W$ can be predicted using the \textit{experimentally }measured maximum curvature in more irregular detonations, for which more non-idealities addressed above are present.  We thus report detonation diffraction experiments in a range of mixtures of varying cellular regularity: $\text{C}_\text{2}\text{H}_\text{6}/\text{3.5O}_\text{2}$, $\text{C}_\text{2}\text{H}_\text{4}/\text{3}\text{O}_\text{2}$ and $\text{C}\text{H}_\text{4}/\text{2}\text{O}_\text{2}$.  These experiments are performed at conditions such that the cellular structure is 3D, i.e., $L>\lambda$ but sufficiently narrow as to not encounter substantial stochasticity in the third dimension.  The critical conditions for diffraction are then compared with the model formulated recently by Radulescu et al.\cite{Radulescu:2021}, using the $D(\kappa)$ limiting conditions obtained experimentally by Xiao and Radulescu \cite{Xiao:2020}.  The model prediction is extended also to other mixtures where experimental maximum front curvature is available \cite{nakayama2013front, Radulescu:2018, xiao:2020dynamics}.

\section{Experiments} \addvspace{10pt}
\subsection{Experimental details} \addvspace{10pt}
Experiments were conducted in a 3.4-m-long thin aluminum channel with an internal cross-section height and width of 0.203 m and 0.019 m, respectively. The schematic illustrating the experimental set-up is shown in Fig.\ \ref{STDI}. The rectangular shock tube comprises three parts, i.e., the detonation initiation section, the propagation section, and the glass-equipped test section for visualization purposes, as detailed elsewhere \cite{xiao:2020dynamics}. The premixed combustible mixture was ignited in the first part by a capacitor discharge.  Inserted mesh wires ensured a detonation was formed in the first section of the tube. Subsequently, detonations travelled in the second propagation part and then entered the third test section. For the present diffraction experiments, a rectangular polyvinyl-chloride plate of 0.683 m in length was placed at the start of the test section. The height of the plate block was 0.165 m, such that the opening height permitting detonations to propagate was 0.038 m, which is twice of the channel width and less than 1/5 of the channel height. Once exiting the reduced opening section, detonations diffracted around the sharp corner into the much larger cross-section area.  The diffraction process was then visualized by utilizing the classical Z-type schlieren technique with the Phantom v1210 camera at 77108 frames per second (about 12.9 $\mu s$ for each interval). The schlieren visualization was implemented with a vertical knife edge utilizing a light source of 360 W, with the exposure time set to 0.44 $\mu$s and the frame resolution kept at $\text{384} \times \text{288}\,\text{px}^\text{2}$. Five 113B27 and three 113B24 piezoelectric PCB pressure sensors were mounted on the top wall of the shock tube for recording pressure signals. The set-up is identical to our previously reported diffraction experiments in H$_2$/O$_2$/Ar detonations \cite{Mevel:2017}.

Three different hydrocarbon-oxygen mixtures were tested, stoichiometric ethane-oxygen ($\text{C}_\text{2}\text{H}_\text{6}/\text{3.5O}_\text{2}$), ethylene-oxygen ($\text{C}_\text{2}\text{H}_\text{4}/\text{3}\text{O}_\text{2}$) and methane-oxygen ($\text{C}\text{H}_\text{4}/\text{2}\text{O}_\text{2}$). Detonations in these mixtures have an irregular cellular structure.  The characteristic structure of these detonations and the dependence of detonation speed on front curvature have been measured in these mixtures by Xiao and Radulescu \cite{xiao:2020dynamics}.   Each mixture was prepared in a separate mixing tank by the method of partial pressures and was then left to mix for more than 24 hours. Once the shock tube was evacuated below the absolute pressure of 80 Pa in every single experiment, the mixture was then introduced into the shock tube through both ends of the channel at the desired initial pressure.  The initial pressure was varied in order to control the mixture sensitivity and identify the critical conditions for diffraction. 

We report the critical diffraction condition in terms of the channel height $W$ normalized with the either $\Delta_i$, the ZND induction zone length or by the cell size $\lambda$.  The induction zone length was calculated using an in-house ZND code \cite{Radulescu1:2018} and the San Diego chemical mechanism.  Corrections for the velocity deficits were incorporated in the calculations by using the experimentally determined detonation velocity.   The cell size for each mixture was taken from the Detonation Database \cite{kaneshige:1997}, fitted to the relation $\lambda=Ap_0^{-B}$, where the fitting parameters A and B for each mixture are presented in table 4 of Xiao and Radulescu \cite{Xiao:2020}.

\subsection{Results} \addvspace{10pt}

Typical examples of the different flow fields obtained near the critical diffraction conditions are shown in Fig.\ \ref{AllExp}.  For all experiments, we evaluated the top-wall mean propagation speed of detonations, before being impacted by the penetration of failure waves, to be in the range of 0.97$\sim$0.99 $D_\text{CJ}$, which is very close to the ideal Chapman-Jouguet (CJ) detonation speed (2223 m/s for ethylene, 2250 m/s  for ethane and  2330 m/s for methane mixtures). It thus suggests that detonations before diffraction can be reasonably assumed to be ideal and losses to the transparent walls are small. 

Figure \ref{AllExp}a shows the unsuccessful transmission of a detonation in $\text{C}_\text{2}\text{H}_\text{4}/\text{3}\text{O}_\text{2}$.  Multiple frames are shown superimposed in the same image.  The detonation originating in the channel of dimension $W/2= 38$ mm, where $W$ is the typical bi-diffraction channel height, is clearly evident to have a cellular structure.  The cellular structure starts being affected by the expansion wave originating from the corner in the fourth frame.  Since the schlieren photographs record density gradients, the demarcation between burned and shocked but non-burned gases is very clear.  This reaction zone is seen to systematically fall behind the lead shock.  In this experiment, the lead shock continuously decays to low speeds and the detonation is quenched by the diffraction process.    

The details of the critical diffraction in a more sensitive mixture obtained at a slightly higher pressure are shown in Fig. \ref{AllExp}b.  At these critical conditions, the detonation wave remains coupled near the top wall.  The cellular structure, although enlarged, remains.  In the region below approximately the corner level, the detonation decouples as for the subcritical case discussed above.  The reformation of a detonation wave proceeds from the growth of a kernel originating near the axis, although the generally irregular structure does not permit to identify the location and mechanism.  Very similar sequence of events have recently been reported in more regular H$_2$/O$_2$/Ar detonations \cite{Radulescu:2021, gallier2017detonation, pintgen:2009}.   In these studies, new detonation heads were preferentially observed slightly off-axis.  Our results show that although the establishment of the detonations with the smallest cells are slightly off-axis, the front near the axis (top wall in our experiments) never fails for successful experiments.  

A similar sequence of events were also observed in $\text{C}_\text{2}\text{H}_\text{6}/\text{3.5O}_\text{2}$ mixture.  Two typical front evolution records bracketing the critical conditions for transmission are shown in Figs. \ref{AllExp}c and d.  The cellular structure in this mixture is more irregular. This is compatible with stability considerations, reflected by the instability parameter, $\chi=\left(t_i/t_r\right) \left(E_a/RT_{VN} \right)$ \cite{radulescu2003propagation}, the product of the ratio of induction and reaction time and the temperature sensitivity of the induction zone).   For the mixtures tested,  $\chi$ takes the value of 17, 45 and 480 for the ethylene, ethane and methane mixtures respectively \cite{Xiao:2020}.

\begin{figure}
\centering
\includegraphics[width=162pt]{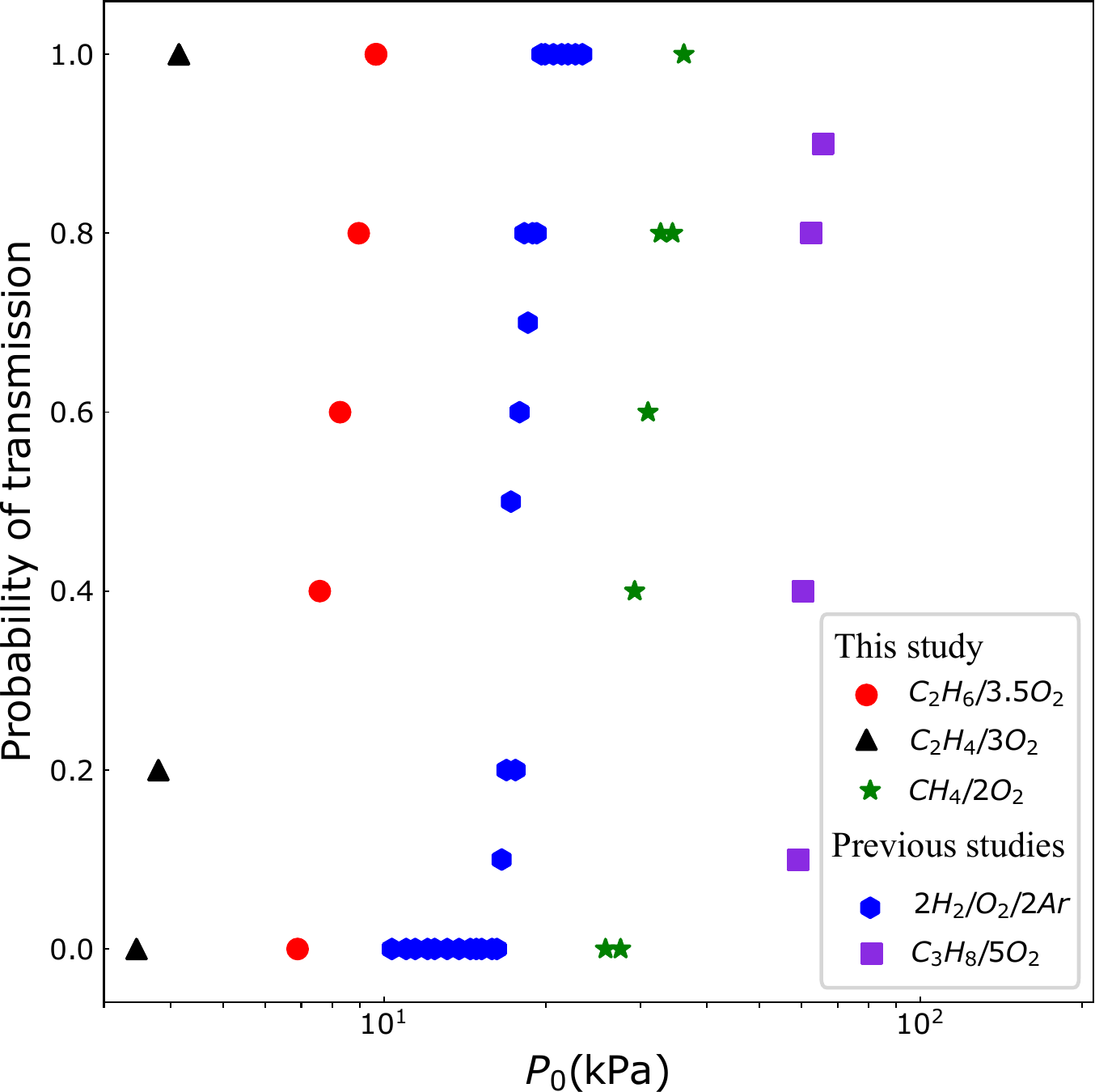}
\caption{Probability of successful transmission of a detonation as a function of initial pressure. The $2\text{H}_\text{2}/\text{O}_\text{2}/\text{2Ar}$ data are from 2D diffraction experiments reported in \cite{Mevel:2017} and $\text{C}_\text{3}\text{H}_\text{8}/5\text{O}_\text{2}$ are 3D diffraction experiments reported in \cite{loiseau:2007}. Diffraction experiments of hydrogen and propane mixtures were repeated 10 times at each pressure. }
\label{Prob}
\end{figure}

The experiments performed in the less sensitive mixture of $\text{C}\text{H}_\text{4}/\text{2}\text{O}_\text{2}$ were not able to produce a successful transmission until $p_0= $ 30 kPa.  However, we only visualized the detonation for subcritical pressures below 26 kPa for safety reasons.   For higher pressures, we used aluminum side plates instead of glass. The successful/unsuccessful transmission of detonation diffraction was identified using pressure transducers. In these experiments the rectangular obstacle inside the shock tube was moved back 0.8 m to have 7 pressure transducers while the diffraction is occurring. The highest pressure tested for successful transmission was 38 kPa. Two typical sub-critical detonation diffraction events are shown in Figs.\ \ref{AllExp}e and f.  In the lower pressure experiment of Fig.\ \ref{AllExp}e, the decoupling of the reaction zone from the diffracting lead shock is clearly evident.  The saw-tooth shape of the reaction contour is the vestigial structure of the original cellular structure.   At the highest pressure where visualization was possible,  shown in Fig.\ \ref{AllExp}f, although the detonation eventually fails, a complex filamentous structure is observed.  

Near criticality, repeated experiments showed that both transmission and failure were observed over a particular range. Repeated experiments were performed at each pressure tested and a probability for transmission was established.  Figure \ref{Prob} provides the probability of detonation transmission for the mixtures tested, along with previous data obtained in previous studies.  The critical pressure for transmission are $8.3 \pm 1.4$ kPa for $\text{C}_\text{2}\text{H}_\text{6}/\text{3.5O}_\text{2}$, $ 3.7 \pm 0.3 $ kPa for $\text{C}_\text{2}\text{H}_\text{4}/\text{3O}_\text{2}$ and $32 \pm 4$ kPa for $\text{C}\text{H}_\text{4}/\text{2O}_\text{2}$ mixtures.  The range of pressures in which both success and failure are possible represents approximately 34\%, 17\% and 25\% of the critical pressures recorded for $\text{C}_\text{2}\text{H}_\text{6}/\text{3.5O}_\text{2}$, $\text{C}_\text{2}\text{H}_\text{4}/\text{3O}_\text{2}$ and $\text{C}\text{H}_\text{4}/\text{2O}_\text{2}$ mixtures, respectively.   The large spread in critical conditions for transmission is in good agreement with the results measured by Loiseau and Higgins \cite{loiseau:2007} for propane-oxygen detonations diffraction in tubes (ranges over 6.7 kPa). Their apparatus consists of a 1.8-m-long tube with a 5-cm-internal-diameter followed by a 40-cm-long cylindrical test chamber with a 19-cm-internal-diameter. The critical pressure of detonation diffraction experiments in $2\text{H}_\text{2}/\text{O}_\text{2}/\text{2Ar}$ mixture spread over 3 kPa as shown in Fig. \ref{Prob}. These experiments were reported by Mevel et al.\ \cite{Mevel:2017} obtained in the same narrow channel used in this study. 

It is more convenient to present the critical diffraction condition in terms of the channel height $W$ normalized with either $\Delta_i$, the ZND induction zone length or by the cell size $\lambda$.   The resulting critical $W_c/\Delta_i$ and $W_c/\lambda$ and given in Table \ref{tab:exp}. Also provided is an estimate of the number of cells across the channel thickness close to the limit, given by $L/\lambda$.  In the methane experiments, $L/\lambda>1$, while approaching unity in the other mixtures.  This suggests that the methane experiments approached the desired 3D cellular structure, while the other mixtures may have been somewhat affected by wall effects. Interestingly, the limit observed in the methane system was precisely half of the $d_c/\lambda = 13$ correlation for tubes as expected from simple geometrical scaling based on wave curvature, while somewhat lower in the other two mixtures. We have not pursued further investigation into these differences, given that cell-size data are usually accurate to within a factor of 2.


\begin{table}\footnotesize
\centering
\caption{Critical diffraction conditions obtained experimentally.}
\resizebox{\columnwidth}{!}{
\begin{tabular}{lcccc}
\hline
 Mixture  &  $P_{0}$ (kPa) & $W_c/\Delta_i$ & $W_c/\lambda$ & $L/\lambda$\\
\hline 
$\text{C}_\text{2}\text{H}_\text{6}/\text{3.5O}_\text{2}$ & 8.3$\pm$1.4  &  170$\pm$28 & 3.5$\pm$0.6 & $\simeq$ 0.7 \\ 	

$\text{C}_\text{2}\text{H}_\text{4}/\text{3}\text{O}_\text{2}$ & 3.7$\pm$0.3 & 150$\pm$13 & 3.8$\pm$0.3 & $\simeq$ 1.0\\

$\text{C}\text{H}_\text{4}/\text{2}\text{O}_\text{2}$& 32$\pm$4& 51$\pm$7 & 6.9$\pm$ 0.9 & $\simeq$ 1.7\\
\hline 
\end{tabular}}
\label{tab:exp}
\end{table}

\begin{figure*}
\centering
\vspace{-0.4 in}
\includegraphics[scale=1]{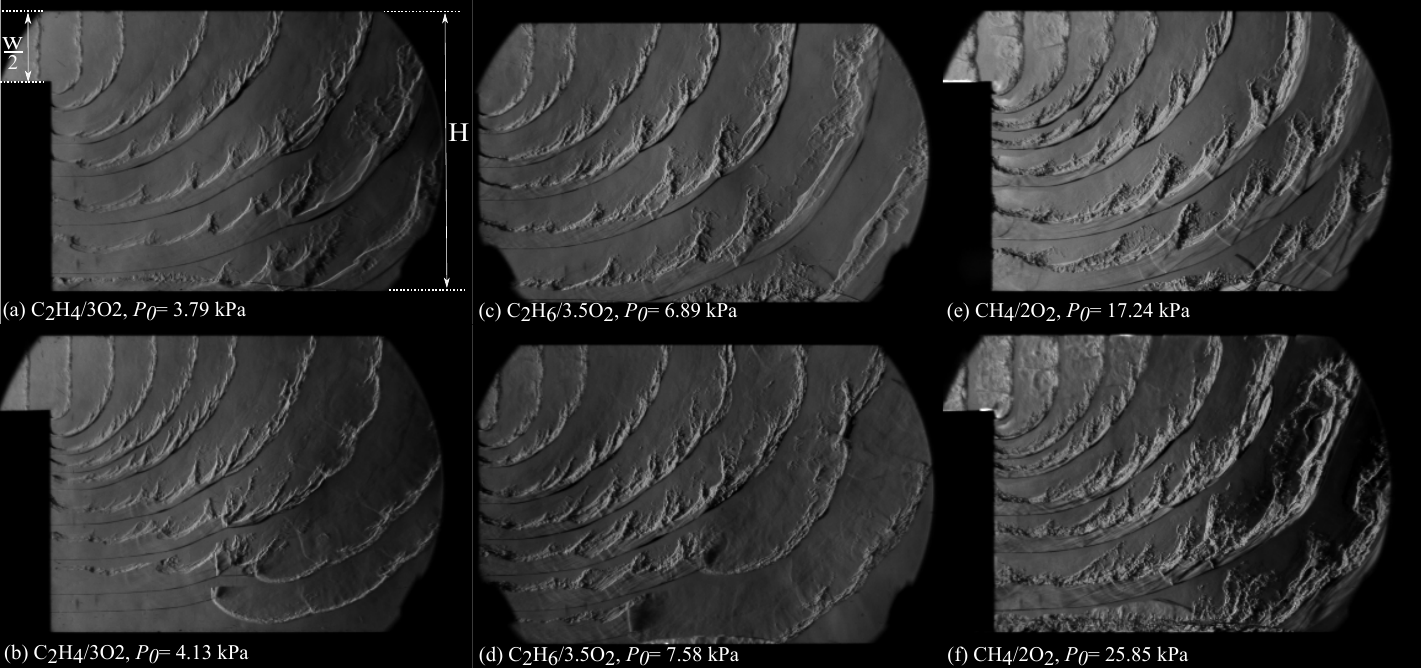}
\vspace{10 pt}
\caption{Composite schlieren images of detonation diffraction in the three mixtures tested at initial temperature 295 K and varying pressures; the distances $H$ and $W/2$ indicated in (a) are 203 mm and 38 mm, respectively. Video animations are provided as supplemental material illustrating the detailed evolution process.}
\label{AllExp}
\end{figure*}
 
\section{Predicted critical slot height $W_c$ from experimental $D(\kappa)$ data} \addvspace{10pt}
As explained in the introduction, we wish to test the prediction of critical diffraction slot height using experimental $D(\kappa)$ data.  Nakayama et al.\ \cite{nakayama2013front} reported that $D(\kappa)$ data for $\text{C}_\text{2}\text{H}_\text{4}/\text{3}\text{O}_\text{2}$, the same mixture as tested in the present study, was characterized by a maximum curvature of $\kappa \lambda\simeq 0.11$, which was found similar to the value in  $\text{2H}_\text{2}/\text{O}_\text{2}$ and $\text{2C}_\text{2}\text{H}_\text{2}/\text{5}\text{O}_\text{2}$/\text{7Ar}.  Xiao and Radulescu \cite{Xiao:2020} determined the $D(\kappa)$ data and the maximum curvature in the mixtures tested in the present study, normalized by the induction zone length.  These data can readily be used to predict the detonation diffraction critical slot width provided a link is found expressing the relation between the slot width $W$ and the maximum frontal curvature of the front.  This link is provided by the study of the subcritical shock diffraction recently formulated by Radulescu et al.\ \cite{Radulescu:2021}.  Since the diffraction of the detonation leads to de-coupling of the reaction zone from the lead shock except for a kernel near the top wall, the curvature at critical conditions is expected to be the curvature imposed by the decoupled shock.  This de-coupled shock evolves as an inert shock.  The problem is approximately self-similar. 

\subsection{The curvature imposed by the decoupled shock}\addvspace{10pt}
In essence, the self-similar shock diffraction can be found in closed form using Whitham's geometrical shock dynamics method \cite{whitham:2011}, provided closure of the exponent $n$ describing the local dynamics of the shock is available \cite{Radulescu:2021}.     The exponent $n$ links the shock speed $D$, its deceleration $\dot{D}$ and its curvature $\kappa$:
\begin{align}
n\equiv-\frac{D^2 \kappa}{\dot{D}} \label{eq:whitham}
\end{align}
The maximum curvature occurs when a transverse signal along the shock reaches the axis of symmetry.  Considering the half height of the bi-diffraction channel is $W/2$, Whitham's geometric shock dynamics yields the following relation between the maximum curvature and the channel height \cite{Radulescu:2021}:
\begin{equation}
{W_c}=\frac{2}{\kappa}\left(\frac{\sqrt{n}}{n+1}\right) \label{eq:WvsK}
\end{equation} 
The exponent $n$ can be modeled with different closures, Whitham's characteristic rule \cite{whitham:2011}, Westcott's reactive model \cite{wescott:2004} and Radulescu's weakly support shock hypothesis, the latter being found to provide the best agreement in the tests of hydrogen detonations \cite{Radulescu:2021}.    We thus wish to test how these closures perform for predicting the dynamics of the decoupled shocks in the current experiments of diffracting detonations.  

In the limit of a strong shock, Whitham's well known characteristic rule \cite{whitham:2011} gives 
\begin{equation}
n=1+\frac{2}{\gamma}+\sqrt{\frac{2\gamma}{\gamma-1}}
\end{equation} 
where $\gamma$ is the ratio of specific heats, assumed constant in the model. In this model, the rear boundary conditions have an insignificant influence on the shock dynamics.  Note that the strong shock assumption is relevant to our study, as the Mach number of detonations is approximately  7.

Wescott, Stewart and Bdzil \cite{wescott:2004} assumed the detonation in a quasi-steady state with an embedded sonic surface.  The exponent $n$ was found to be:
 \begin{equation}
n=3\frac{\gamma+1}{\gamma}
\end{equation} 

Radulescu et al. \cite{Radulescu:2021} proposed a weakly supported shock model in which $(1/ \dot{D}) \partial u/ \partial t \ll 1$.  This yielded  $n$:
\begin{equation}
n=2\frac{\gamma+1}{\gamma} \label{eq:nfromR}
\end{equation} 

The resulting value of modeling exponent $n$ from the three models (weak support (R), Whitham (W) and Wescott, Stewart and Bdzil (WSB) are shown in Table \ref{n}, for the critical pressures of diffraction experiments.  The ratio of specific heats was evaluated behind the shock.
\begin{table}\small
\centering
\caption{Summary of the modelling exponent $n$. VN refers to the von Neumann state.}
\resizebox{\columnwidth}{!}{
\begin{tabular}{lccccc}
\hline
 Mixture  &  $P_{0}$ (kPa)  & $\gamma_\text{ VN}$& $n_\text{R}$&  $n_\text{W}$& $n_\text{WSB}$ \\ 
\hline 
$\text{C}_\text{2}\text{H}_\text{6}/\text{3.5O}_\text{2}$ & 8.3$\pm$1.4&1.15  &3.74  & 6.64&5.60 \\ 

$\text{C}_\text{2}\text{H}_\text{4}/\text{3}\text{O}_\text{2}$ & 3.7$\pm$0.3&1.17  &3.71 &6.42 &5.56 \\

$\text{C}\text{H}_\text{4}/\text{2}\text{O}_\text{2}$& 32$\pm4$ &1.17 &  3.71& 6.43&5.56 \\
\hline
\end{tabular}}
\label{n}
\end{table}

\begin{figure}[h]
\centering
\includegraphics[width=192pt]{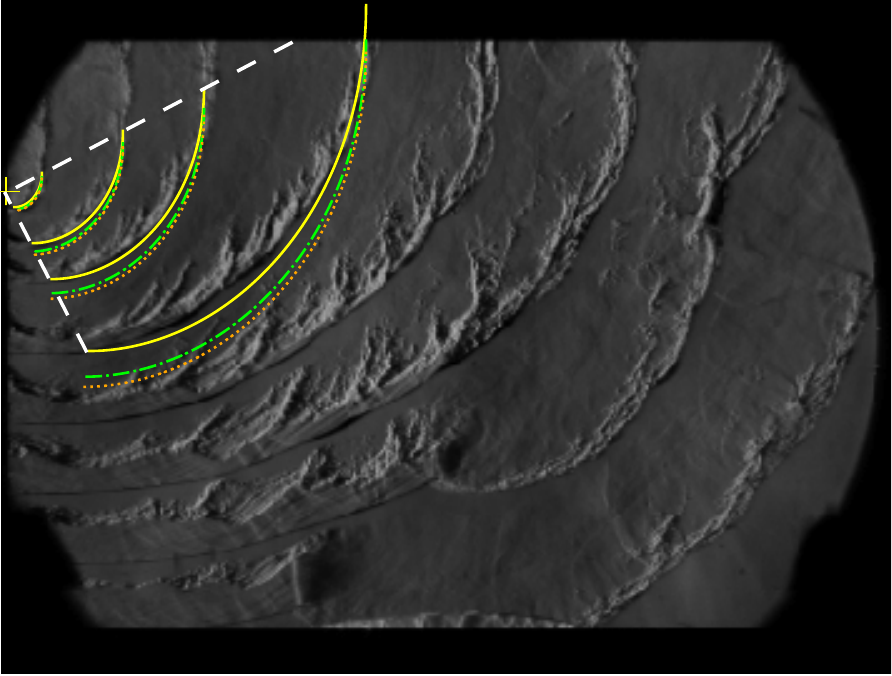}
\caption{Composite schlieren images of detonation diffraction in the critical diffraction of $\text{C}_\text{2}\text{H}_\text{6}/\text{3.5O}_\text{2}$ mixture at initial temperature and pressure 295K and 7.58 kPa.}
\label{curve1}
\end{figure}

\begin{figure}[h!]
\centering
\includegraphics[width=192pt]{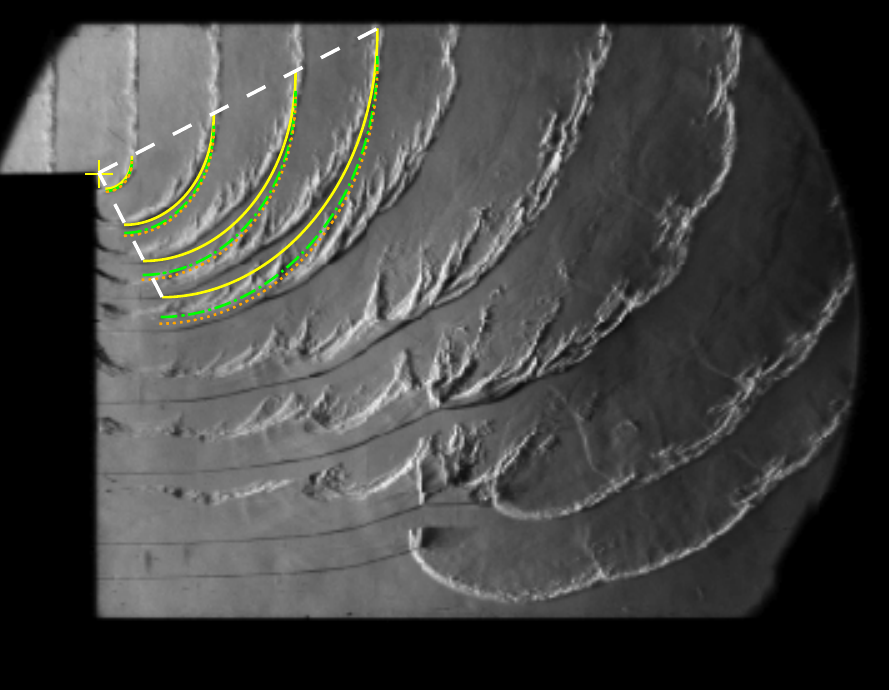}
\caption{Composite schlieren images of detonation diffraction in the critical diffraction of $\text{C}_\text{2}\text{H}_\text{4}/\text{3}\text{O}_\text{2}$ mixture at initial temperature and pressure 295K and 4.13 kPa.}
\label{curve2}
\end{figure}

\begin{figure}[h!]
\centering
\includegraphics[width=192pt]{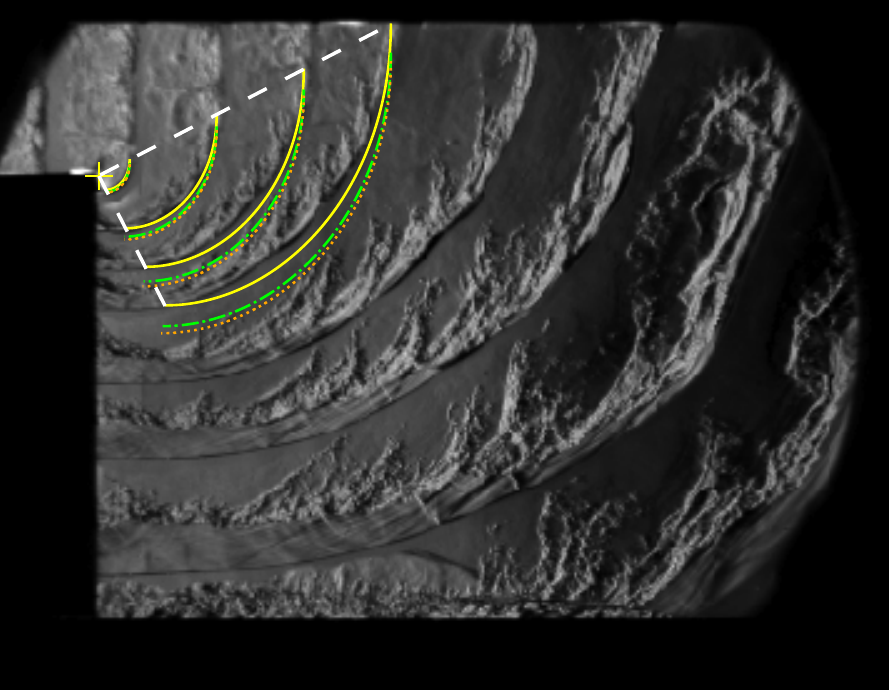}
\caption{Composite schlieren images of detonation diffraction in sub-critical diffraction of $\text{C}\text{H}_\text{4}/\text{2}\text{O}_\text{2}$  mixture at initial temperature and pressure 295K and 25.85 kPa.}
\label{curve3}
\end{figure}

The performance of the three models discussed above in predicting the shock evolution is shown in Figs.\ \ref{curve1}, \ref{curve2} and \ref{curve3} for critically transmitted detonations in $\text{C}_\text{2}\text{H}_\text{6}/\text{3.5O}_\text{2}$, $\text{C}_\text{2}\text{H}_\text{4}/\text{3}\text{O}_\text{2}$ and $\text{C}\text{H}_\text{4}/\text{2}\text{O}_\text{2}$.  In a coordinate system centered at the corner, the shock shape is given by Whitham's self-similar solution: 
\begin{align}
\frac{X}{D_{CJ} t} = \sqrt{ \frac{n+1}{n} } \exp\left( \frac{\theta}{\sqrt{n}} \right) \sin \left( \eta-\theta \right)\\
\frac{Y}{D_{CJ} t} = \sqrt{ \frac{n+1}{n} } \exp\left( \frac{\theta}{\sqrt{n}} \right) \cos \left( \eta-\theta \right)
\end{align}
where $\eta$ is given by $\tan \eta = \sqrt{n}$ and $\theta$ is the angle of the unit normal to the shock surface with the $x$-axis. In all the figures, the solid yellow lines correspond to the weakly supported shock model, the dash-dotted green lines to the WSB model, and the dotted orange lines to Whitham’s inert shock model, at the sequential frames of experiments. The dashed white lines are the characteristic lines which show the propagation of the signal from the corner for the weakly supported shock model of Radulescu. Accordingly, the region where the model is valid is between the indicated characteristic lines.

In all of the diffraction experiments, the weakly supported self-similar model is predicting the curvature of the diffracted shock very well from early to later times of diffraction with maximum error of curvature 20\% for $\text{C}_\text{2}\text{H}_\text{6}/\text{3.5O}_\text{2}$ and $\text{C}\text{H}_\text{4}/\text{2}\text{O}_\text{2}$, and 15\% for $\text{C}_\text{2}\text{H}_\text{4}/\text{3}\text{O}_\text{2}$ mixtures. These results show that the shock support model of Radulescu offers an improvement over the other two models, which under-estimate somewhat the shock wave curvature.

\subsection{Prediction of critical channel height} 
\begin{figure}[h!]
\centering
\includegraphics[width=192pt]{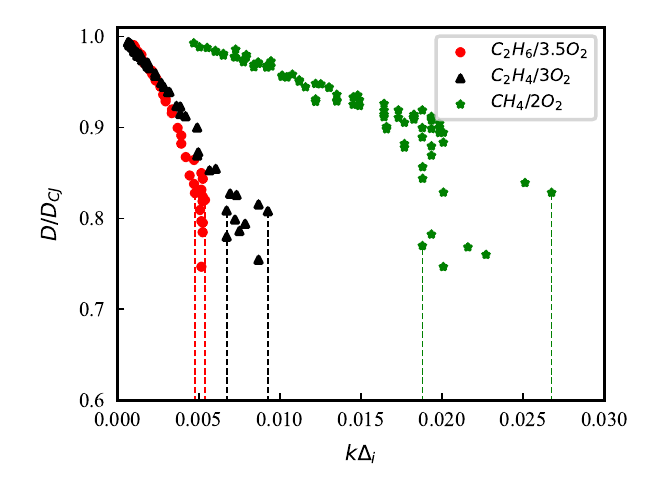}
\caption{The non-dimensional $D-\kappa$ characteristic relationships obtained experimentally, adapted from \cite{Xiao:2020}.}
\label{ramp}
\end{figure}

\addvspace{10pt}
Given the good agreement of the weakly supported model to predict the shock wave evolution, we substituted the value of $n$ provided by Eq. \eqref{eq:nfromR} in Eq. \eqref{eq:WvsK} to close the model for the critical channel height for successful detonation diffraction, in terms of the maximum curvature.  This can also be re-written as:
\begin{equation}
\frac{W_c}{\Delta_i}=2\frac{\sqrt{n}}{(n+1)\kappa\Delta_i} \label{eq:WvsK2}
\end{equation}
The maximum curvature required for closure is obtained from the $D/D_{CJ}=f\left(\kappa\Delta_i\right)$ empirical data measured by Xiao and Radulescu\ in the exponentially diverging ramp in the mixtures tested in the present study. These data are reproduced in Fig.\ \ref{ramp} \cite{Xiao:2020}. Their results show a good scaling of detonation deficit with the non-dimensional curvature  $\kappa\Delta_i$.  With increasing curvature, the detonation is slower and there is a maximum curvature that can sustain a steadily propagating cellular detonation.  We use this maximum curvature $\kappa\Delta_i$, which is in the range of 0.0047-0.0054 for $\text{C}_\text{2}\text{H}_\text{6}/\text{3.5O}_\text{2}$, 0.0067-0.0093 for $\text{C}_\text{2}\text{H}_\text{4}/\text{3}\text{O}_\text{2}$ and 0.0185-0.027 for $\text{C}\text{H}_\text{4}/\text{2}\text{O}_\text{2}$.   The uncertainty ranges are shown as vertical dashed lines in Fig.\ \ref{ramp}. These provided $D-\kappa$ data for cellular detonations near failure shown in Fig.\ \ref{ramp} have certainly a measure of stochasticity which is linked with the presence of fluctuations and requires a stochastic approach with very large datasets. Future work needs to establish whether this stochasticity in the $D-\kappa$ limits is sufficient to infer the diffraction stochasticity or not.

The critical values obtained for the $W_c/\Delta_i$ using \eqref{eq:WvsK2} and the data for the maximum curvatures outlined above, along with their confidence levels based, are tabulated in Table \ref{tab:model}.  The ratio of specific heats required were calculated at the Von Neumann state for each mixture at the critical diffraction conditions.  These predicted values of $W_c/\Delta_i$ can now be compared with the experimental values. Note that the results were corrected to account for the slight lengthening of the induction zone length due to the losses by accounting for the velocity deficits.

\begin{table*}\footnotesize
\centering
\caption{Summary of diffraction experiments and models prediction. Exp: Experiment, R: Radulescu (weakly supported shock model),  WSB: Wescott, Bdzil and Stewart's model and W: Witham's model.}
\begin{tabular}{lccccc}
\hline
 Mixture  &   $W_c/\Delta_i$, Exp & $W_c/\Delta_i$, Model, R & $W_c/\Delta_i$,WSB  &	$W_c/\Delta_i$, W	& $W_c/\Delta_i$, ZND\\
\hline 
$\text{C}_\text{2}\text{H}_\text{6}/\text{3.5O}_\text{2}$ &  170$\pm$28 & 161$\pm$12 & 141$\pm9$	 & 	133$\pm9$	&	 593 \\ 	

$\text{C}_\text{2}\text{H}_\text{4}/\text{3}\text{O}_\text{2}$ & 150$\pm$13 & 102$\pm$15 & 89$\pm12$ &    85$\pm12$	&	 300\\

$\text{C}\text{H}_\text{4}/\text{2}\text{O}_\text{2}$& 51$\pm$7 &36$\pm$ 6&  	31$\pm$5& 	30$\pm5$	&	767\\
\hline 
\end{tabular}
\label{tab:model}
\end{table*}

The predictions for critical slot width $W_c$ tabulated in Table \ref{tab:model} are in general good agreement with the experimental value.   Among the predictions made with different exponents $n$, the model of Radulescu provides the closest agreement with experiment.  For the C$_2$H$_6$ mixture, the prediction is in excellent agreement with the experiment with less than 5 \% discrepancy.  For the C$_2$H$_4$ and CH$_4$ mixtures, the model under-predicts the experimental limit by 32\% and 38\% respectively.  When considering the uncertaities due to the noted stochasticities, the agreement is in fact much better. We can thus conclude that the model is successful in predicting the critical failure conditions.  The paradigm tested, mainly, can the diffraction of cellular detonations be predicted by a $D(\kappa)$ law for a unit cell, is thus successful.

It is also of interest to test the predictability of the detonation limits using the maximum curvature predicted by the ZND model, as opposed to the experimental value.  This is provided in last column of Table \ref{tab:model}, obtained from Eq. 14 in \cite{Radulescu:2021}. As expected, this model significantly over-predicts the limits,  by more than an order of magnitude for the methane system, as expected from our previous work \cite{Xiao:2020}. This highlights again the need to construct a model for detonations that takes into account the cellular processes and how they enhance the propagation mechanism.\\

\subsection{Failure prediction using other experimentally measured $D-\kappa$ data} \addvspace{10pt}

Nakayama et al.\ \cite{nakayama2013front} reported that the $D(\kappa)$ curves for stable detonation propagation in $\text{C}_\text{2}\text{H}_\text{4}/\text{3}\text{O}_\text{2}$, $\text{2H}_\text{2}/\text{O}_\text{2}$ and $\text{2C}_\text{2}\text{H}_\text{2}/\text{5}\text{O}_\text{2}$/\text{7Ar} were identical when scaled with the cell size.  The maximum curvature of $\kappa \lambda\simeq 0.11$ was found.  Re-writing \eqref{eq:WvsK2} normalized by the cell size as: 
\begin{equation}
\frac{W_c}{\lambda}=2\frac{\sqrt{n}}{(n+1)\kappa\lambda} \label{eq:WvsK3}
\end{equation}
and using the exponent $n$ of 3.71, it yields a critical slot width of $W_c/\lambda \approx 7$.  This value is slightly over-predicting the limits found in the current study, in the range of 3-7.  Given that Nakayama et al.\ only report stable propagation, i.e., devoid of the typical non-steadiness observed near the limits, it means that the real maximum curvature is somewhat larger and the limit $W_c/\lambda$ is smaller than 7.  Nevertheless, the limit reported by them is also compatible with the diffraction in tubes, where one would expect that $d_c/\lambda$ to be twice as for channels.  This further supports the idea that critical diffraction of cellular detonations can be predicted by $D(\kappa)$ data for cellular detonations. 

	Datasets for $D(\kappa)$ were obtained in other mixtures by Radulescu and Borzou \cite{Radulescu:2018} and Xiao and Radulescu \cite{xiao:2020dynamics} in mixtures of $\text{C}_\text{3}\text{H}_\text{8}/\text{5O}_\text{2}$, $\text{2C}_\text{2}\text{H}_\text{2}/\text{5O}_\text{2}/\text{21Ar}$,  $\text{2H}_\text{2}/\text{O}_\text{2}/\text{2Ar}$, $\text{2H}_\text{2}/\text{O}_\text{2}/\text{3Ar}$ and $\text{2H}_\text{2}/\text{O}_\text{2}/\text{7Ar}$ using the exponential horn technique.  Xiao and Radulescu \cite{Xiao:2020} reduced these data in terms of $\kappa\lambda$ (figure 18 in their paper).  The limiting value of curvature was found to be in the range of 0.2 to 0.6, the larger values for more regular H$_2$/O$_2$/Ar system.  This yields, according to \eqref{eq:WvsK3}, limiting diffraction values of 2 to 4.  This agrees well with the experimental value of 3 to 6 for the more irregular detonations but underpredicts the limits generally accepted for more regular detonations.  Empirically, it thus appears that the DSD treatment of cellular detonations, whereby diffraction can be predicted by experimentally measured $D(\kappa)$ datasets work better for more irregular systems.
	
	At this point, we can only speculate why a quasi-steady $D(\kappa)$ response is more adequate for irregular detonations,  based on the physics of the propagation of regular and irregular detonations.  Regular detonations behave similarly to laminar ZND-detonations where the ignition mechanism is by adiabatic compression. As the limits are approached, non-steady effects can no longer be neglected and the structure no longer responds as in quasi-steady state.  On the other hand, the ignition mechanism of irregular detonations is by a combination of adiabatic compression near the front and diffusion assisted burning of pockets at the later stages.  The latter does not have the strong coupling to non-steady gasdynamic effects, and a more quasi-steady response is expected.   Since the thin front responds much faster than the thick main reaction zone, one would expect the entire structure to remain in quasi-steady state on the longer time scales associated with the entire reaction zone structure.  Empirically, we also find that the quasi-steady prediction to work better for irregular mixtures, where this particular reaction structure is more prevalent.  Clearly, however, these transient effects need to be carefully studied in the future for cellular detonations.  It is expected that these transient effects influence not only the limits in diffraction phenomena but also the methods of measurement of the steady responses to begin with, where entrance effects need to be carefully evaluated.

\section{Closing remarks} \addvspace{10pt}
In the present study, we have showed that the critical diffraction of irregular detonations can be very well predicted from an experimentally determined $D(\kappa)$ law.  The prediction error of less than 5\% was observed for $\text{C}_\text{2}\text{H}_\text{6}/\text{3.5O}_\text{2}$, while the limits were underpredicted by 32\% for $\text{C}_\text{2}\text{H}_\text{4}/\text{3}\text{O}_\text{2}$, and 38\% for $\text{C}\text{H}_\text{4}/\text{2}\text{O}_\text{2}$ mixtures. Given the inherent stochasticity near failure in the diffraction experiments and available $D(\kappa)$ of approximately 20 to 30\% for each, the agreement is actually much better.

This generally good agreement supports the view that the detonation diffraction critical conditions are associated with a maximum rate of frontal stretch, or curvature \cite{lee1996, wescott:2004}.  More generally, the present study suggests that the physics of propagating detonations can be encapsulated in the problem of a single unit cell propagating in a diverging ray tube.  It highlights the necessity for models to be developed for this canonical problem, that account for cellular instabilities. 

\acknowledgement{Acknowledgments} \addvspace{10pt}
M.I. Radulescu acknowledges the financial support provided by the Natural Sciences and Engineering Research Council of Canada (NSERC) through the Discovery Grant ”Predictability of detonation wave dynamics in gases: experiment and model development”. 

\acknowledgement{Supplementary material} \addvspace{10pt}
Supplementary material associated with this article can be found in the online version.


 \footnotesize
 \baselineskip 9pt


\bibliographystyle{pci}
\bibliography{Zangene-Proci}

\newpage

\small
\baselineskip 10pt



\end{document}